\documentclass[a4paper,12pt]{article}
\usepackage{a4wide}
\usepackage[top=2.5cm, bottom=2.5cm, left=2.5cm, right=2.5cm]{geometry}
\usepackage[utf8]{inputenc}
\usepackage{graphicx}
\usepackage{amsmath}
\usepackage{hyperref}
\usepackage{url}
\usepackage{color}
\usepackage{cite}

\def\be{\begin{equation}}
\def\ee{\end{equation}}
\def\bea{\begin{eqnarray}}
\def\eea{\end{eqnarray}}

\begin{document}
\title{\textbf{Hunting the elusive $X17$\\[0.25cm] in CE$\nu$NS at the ESS}}
\author
{
J. Cederk{\"a}ll$^{1}$\footnote{E-mail:  \texttt{joakim.cederkall@fysik.lu.se}},
~Y. Hi\c{c}y{\i}lmaz$^{2,4}$\footnote{E-mail:  \texttt{yasarhicyilmaz@balikesir.edu.tr;~y.hicyilmaz@soton.ac.uk}},
~E. Lytken$^{1}$\footnote{E-mail:  \texttt{else.lytken@fysik.lu.se}},
~S. Moretti$^{3,4}$\footnote{E-mail:  \texttt{stefano.moretti@physics.uu.se;~s.moretti@soton.ac.uk}},
~J. Rathsman$^{1}$\footnote{E-mail:  \texttt{johan.rathsman@fysik.lu.se}},\\
\\%
{\small\it $^{1}$ Department of Physics,
Lund University,} 
\\%
{\small\it  221 00 Lund, Sweden}
\\%
{\small\it $^{2}$ Department of Physics, Bal{\i}kesir University,} 
\\%
{\small\it   TR10145, Bal{\i}kesir, Turkey}\\
{\small\it $^{3}$ Department of Physics and Astronomy, Uppsala University,} 
\\%
{\small\it Box  516,  751 20, Uppsala, Sweden  }\\
{\small\it $^{4}$ School of Physics and Astronomy, University of Southampton} 
\\%
{\small\it Highfield, Southampton SO17 1BJ, United Kingdom  }\\
}

\maketitle

\begin{abstract}
The so-called $X17$ particle has been proposed in order to explain a
very significant resonant behaviour (in both the angular separation and invariant mass) of $e^+e^-$ pairs produced  during a nuclear transition of 
excited $^8$Be, $^4$He and $^{12}$C nuclei. Fits to the corresponding data point, as most probable explanation, to a spin-1 object, which is protophobic and has a mass of approximately 16.7 MeV, which then makes the $X17$ potentially observable in
Coherent Elastic neutrino ($\nu$) Nucleus Scattering (CE$\nu$NS) at the 
European Spallation Source
(ESS). By adopting as theoretical framework a minimal extension of the Standard Model (SM) with a generic $U(1)'$  gauge group mixing with the hypercharge one of the latter,  which can naturally accommodate the $X17$ state compliant with all available measurements from a variety of experiments, we predict that  CE$\nu$NS at the ESS will constitute an effective means to probe this hypothesis, even after allowing for the inevitable systematics associated to the performance of the planned detectors therein.
 
\end{abstract}

\clearpage
\section{Introduction}\label{Sect1}

While conventional wisdom assumes that  physics Beyond  the Standard Model (BSM)  would naturally manifest itself at high energies, potentially beyond the  reach of current laboratory experiments, like the forefront Large Hadron Collider (LHC),
it must be noted that  one of the most persistent data anomalies to date with respect to  SM predictions has  actually been established at a (very) low energy facility. We are referring here to
the ATOMKI experiment \cite{Gulyas:2015mia}, which studied over the years the trajectory paths of particles produced during nuclear transitions of $^8{\rm Be}$, $^4{\rm He}$ and $^{12}{\rm C}$ 
\cite{Krasznahorkay:2015iga,Sas:2022pgm,Krasznahorkay:2017gwn,Krasznahorkay:2017bwh,Krasznahorkay:2017qfd,Krasznahorkay:2018snd,
Krasznahorkay:2019lyl,Krasznahorkay:2021joi,Krasznahorkay:2022pxs}.
Herein, a significant enhancement, of well more than 5$\sigma$ (when combined across all elements), was observed, in the form of 
excess $e^+e^-$ decays emerging at consistently similar opening angles with an invariant mass of approximately 16.7 MeV. PADME also has sensitivity to this mass and have recently posted their results in Ref.~\cite{Bossi:2025ptv},
claiming a $2\sigma$  excess corresponding to the mass indicated by the ATOMKI experiment. As for the MEG-II experiment \cite{MEGII:2024urz}, their data disfavour the ATOMKI hypothesis and set an upper limit on the decay rate of a potential new particle into electron-positron pairs\footnote{See {https://cerncourier.com/a/mixed-signals-from-x17/ for a recent popularised account of the ATOMKI anomaly, including a concise review of its possible explanations}.}.

While waiting for a resolution to such a conundrum, we assume here that these effects can be interpreted as due to a new object, so-called $X17$, a spin-1 boson that could well be the force carrier of a postulated fifth force, possibly connected with Dark Matter (DM).
In fact,  Ref.~\cite{Feng:2016jff} (see Refs. \cite{Feng:2016ysn,Feng:2020mbt,Nomura:2020kcw,Seto:2020jal,Kozaczuk:2016nma,DelleRose:2018pgm,DelleRose:2019hnc} for additional explanations)  showed that, in the case of a pure vector coupling, the new boson should be protophobic, since its allowed couplings to nucleons are strongly constrained by the NA48/2 experiment~\cite{NA482:2015wmo}. 
Further phenomenological studies~\cite{Barducci:2022lqd,Denton:2023gat} demonstrated that, due to conflicts with the non-observation of deviations from the SM in neutrino scattering experiments,  the scenarios including a pure vector  mediator  are less favourable,  while an axial-vector state appears as the most promising candidate to simultaneously explain all the anomalous nuclear decays reported by the ATOMKI collaboration~\cite{Alves:2023ree}.

Furthermore, several studies have been conducted to investigate the effects of such a light $X17$  also on the anomalous magnetic moment of the electron ($a_e$) and muon ($a_\mu$)  as well as at the LHC (including flavour observables such as $R_{K^{(*)}}$)~\cite{Nomura:2020kcw,Seto:2020jal,DelleRose:2017xil,DelleRose:2019ukt,Barman:2021yaz,Bodas:2021fsy,Fayet:2020bmb,Hati:2020fzp}, all pointing  to the $X17$ consistency with all such datasets.  This is all the more intruiguing as attempts to interpret the ATOMKI results in terms of background effects have demonstrated that  nuclear physics or QCD cannot lead to a satisfactory  explanation~\cite{Zhang:2017zap,Koch:2020ouk,Chen:2020arr,Aleksejevs:2021zjw,Kubarovsky:2022zxm,Hayes:2021hin,Viviani:2021stx,Hicyilmaz:2022owb}.  
A minimal theoretical approach embedding the $X17$ is given by a family-dependent $U(1)$ extension of the SM, which breaking introduces a new light vector boson, effectively a $Z'$, which can indeed be interpreted as the $X17$, as first introduced in Ref.~\cite{DelleRose:2018eic}. This naturally  allows for  axial-vector couplings, depending on the gauge quantum numbers involved. Furthermore, in this framework, the Yukawa interactions are suitably modified by higher-dimensional operators~\cite{Pulice:2019xel}. 

Notice, in particular, that this scenario does not prevent $Z'$ couplings to neutrinos, so that it can also lead to Non-Standard Interactions (NSIs) of the latter affecting neutrino flavour ratios  in matter \cite{Proceedings:2019qno}. Therefore, the ensuing experimental constraints on NSIs from neutrino oscillations can be applied to restrict the family dependent (non-universal) couplings of the new boson to SM fermions, as recently done (using data from TEXONO~\cite{TEXONO:2009knm,Bilmis:2015lja}
and IceCube~\cite{IceCubeCollaboration:2021euf}) in Ref.~\cite{Enberg:2024ofo}.

Given the possibility of such a $Z'$ interacting with neutrinos, in the spirit of Ref.~\cite{Chattaraj:2025rtj}, we investigate here  the possibilities to discover this potential carrier of a fifth force 
in Coherent Elastic neutrino ($\nu$) Nucleus Scattering (CE$\nu$NS) at the European Spallation Source(ESS), building on
the work done in \cite{Enberg:2024ofo}. Specifically, we will show that the aforementioned $X17$  will not only modify the CE$\nu$NS rate which could be observed in an experiment at the ESS but also the shape of the nuclear recoil spectrum (see Ref.~\cite{Procs} for an initial account of this).

The plan of the paper is  as follows. In the next section we introduce our BSM  framework while in the following one we describe the theory of CE$\nu$NS at the ESS. We then present Sect.~\ref{Sect4}, where we make some considerations about the experimental setup required at the ESS to see evidence of the $X17$ in CE$\nu$NS. Sect.~\ref{Sect5} assesses the discovery reach of the latter. We finally conclude in Sect.~\ref{Sect6}.

\section{The light $Z^\prime$ model}\label{Sect2}

In this work, we study a minimal extension of the SM with a generic $U(1)'$  gauge group which mixes with the hypercharge one of the SM,  $U(1)_Y$. Such a model has kinetic terms in its Lagrangian given by
\begin{equation}
\label{eq:KineticL}
\mathcal{L}_\mathrm{Kin}^{U(1)'} = - \frac{1}{4} \hat F_{\mu\nu} \hat F^{\mu\nu}  - \frac{1}{4} \hat F'_{\mu\nu} \hat F^{'\mu\nu} - \frac{\eta}{2} \hat F'_{\mu\nu} \hat F^{\mu\nu},
\end{equation}
where the field strengths $  \hat F_{\mu\nu} $  and $ \hat  F'_{\mu\nu} $ correspond to the gauge fields $  \hat B_\mu $ and $ \hat  B'_\mu$ of $U(1)_Y$ and $U(1)'$, respectively, while the parameter $\eta$ quantifies the kinetic mixing between these Abelian symmetries. Transforming the fields to remove the kinetic mixing by
\begin{eqnarray}
\hat{B}_{\mu} &=& B_{\mu} - \dfrac{\eta}{\sqrt{1-\eta^2}}B'_{\mu} \\
\hat{B}'_{\mu} &=& \frac{1}{\sqrt{1-\eta^2}}B'_{\mu}	\, ,
\label{eq:kinZZprotation}
\end{eqnarray}
gives the gauge covariant derivative as
\begin{equation}
\label{CovDer}
{\cal D}_\mu = \partial_\mu + \dots + i g_1 Y B_\mu + i (\tilde{g} Y + g' Q') B'_\mu, 
\end{equation}
where $Y$ represents the hypercharge with coupling $g_1$ while $Q'$ denotes the $U(1)'$ charge with associated coupling $g'=\hat g'/\sqrt{1-\eta^2}$, where $\hat g'$ is the original $U(1)'$ coupling before the transformation, and $\tilde{g}= -\eta g_1/\sqrt{1-\eta^2}$ characterises the strength of the gauge mixing between $ U(1)_{Y} $ and $ U(1)' $.

The $U(1)'$ symmetry is spontaneously broken by a new SM singlet scalar $ \chi $, 
with 
Vacuum Expectation Values (VEVs)  $\langle\chi\rangle =  v'/\sqrt{2}$ in addition to the SM VEV $v$ as will be discussed in more detail below.
This breaking generates a mass term for an additional vector boson. The neutral gauge boson mass eigenstates are obtained through the rotation:
\bea
\left( \begin{array}{c} B^\mu \\ W_3^\mu \\ B'^\mu \end{array} \right) = \left( \begin{array}{ccc} 
	\cos \theta_W & - \sin \theta_W \cos \theta' & \sin \theta_W \sin \theta' \\
	\sin \theta_W & \cos \theta_W \cos \theta' & - \cos \theta_W \sin \theta' \\
	0 & \sin \theta'  & \cos \theta'  
\end{array} \right)
\left( \begin{array}{c} A^\mu \\ Z^\mu \\ Z'^\mu \end{array} \right),
\eea
where $\theta_W$ is the Weinberg angle and $\theta'$ is $Z-Z'$ mixing angle \cite{Coriano:2015sea}:
\bea \label{ThetaPrime}
\tan 2 \theta' = \frac{2 g_H g_Z }{ g_H^2 + ( 2 Q'_\chi g' \, v'/v )^2 - g_Z^2} \,,
\eea
with $Q'_\chi$ being the $U(1)'$  charge of the $\chi$-field, $ g_H = \tilde{g} + 2 g' Q'_H $ and $g_Z=\sqrt{g_2^2 + g_1^2} $ is the Electro-Weak (EW) coupling. The corresponding $Z$ and $Z'$ masses are then given by
\begin{equation}
    m_{Z,Z'}^2=   \frac{v^2}{4} g_Z^2 \left(1+\frac{g_H^2+(2Q'_\chi g' \, v'/v)^2}{g_Z^2} \mp \frac{g_H}{\sin2\theta' g_Z}\right) \, .
\end{equation}
The LEP precision measurements constrain the mixing $\theta'$ to be such that $|\theta'| \ll 10^{-3}$ \cite{DELPHI:1994ufk,Erler:2009jh}. In the ensuing small $\theta'$ approximation (treating the difference between $\sin 2 \theta'$ and $\tan 2 \theta'$ carefully), the gauge boson masses become 
\bea \label{ZZpMassesSmallTh}
m_Z^2 \simeq  \frac{v^2}{4} g_Z^2
\,, \qquad 
m_{Z'}^2 \simeq ( Q'_\chi g' \, v')^2
\eea
while the mixing angle simplifies to
\bea \label{ThetaPrimeMasses}
\theta' \simeq -\frac{g_H}{g_Z}  \,.
\eea
where $g_H^2 \ll g_Z^2$ and $(g' v')^2 \ll (g_Z v)^2$  have also been assumed.
For $g' \sim {\cal O}(10^{-4} - 10^{-5})$, $M_{Z'}$ becomes light (${\cal O}(10)$ MeV) when $ v' \approx {\cal O}(100 -1000) $ GeV,  then potentially offering a candidate to address the ATOMKI anomaly.

The $  Z' $ interaction Lagrangian with SM fermions is
\begin{eqnarray}
\label{eq:NeuCurLag}
\mathcal{L}^\mathrm{Z'} &=& \bar q \gamma^\mu \left( C^{qq'}_{L} P_L + C^{qq'}_{R} P_R \right) q' Z'_\mu + \bar \nu_l \gamma^\mu \left( C^{ll'}_{L} P_L \right) \nu_{l'} Z'_\mu \nonumber \\ 
&+&\bar l \gamma^\mu \left( C^{ll'}_{L} P_L + C^{ll'}_{R} P_R \right) l' Z'_\mu,
\end{eqnarray}
where $ q^{(\prime)}$,  $l^{(\prime)}$  and $  \nu_{l^{(\prime)} }$ refer to up-type/down-type quarks, charged leptons and their neutrinos while $ C^{XX}_{L} $ and $ C^{XX}_{R} $ are Left ($L$) and Right ($R$) handed couplings with $P_L$ and $P_R$ the corresponding projection operators $\frac{1\mp\gamma^5}{2}$, respectively. In our model, there are no flavour-violating (non-diagonal) coupling terms for the quark and lepton sectors. In such a flavour-conserving scenario, the diagonal couplings are
\begin{eqnarray}
	C^{ff}_{L} \!&\!=\!&\!  - g_Z \sin \theta' \left( T^3_f - \sin^2 \theta _W Q_f \right) \!+\! ( \tilde g Y_{f, L} \!+\! g' Q'_{f, L})  \cos \theta'\!, \label{CffL}  \\
	C^{ff}_{R} &=&  g_Z \sin^2 \theta_W \sin\theta' Q_f + ( \tilde g Y_{f, R} + g' Q'_{f, R}) \, \cos \theta ', \label{CffR}
	\end{eqnarray}
where $T_f ^3$ and $Q_f$ are the fermion's weak isospin and electric charge, respectively, with $Y_{f,L/R}$ and $Q'_{f,L/R}$ being  the $U(1)_Y$ and $U(1)'$ charges.

From these left- and the right-handed couplings defined via Eqs.~(\ref{CffL}) and (\ref{CffR}) we can then get the vector and axial couplings  as
\bea \label{VAcouplings}
C_{f, V} = \frac{C^{ff}_{R} + C^{ff}_{L}}{2} \,, \quad C_{f, A} = \frac{C^{ff}_{R} - C^{ff}_{L}}{2} .
\eea
 In the limit of small gauge coupling and mixing, $ g^\prime, \tilde{g} \ll 1$, the vector and axial couplings become~\cite{DelleRose:2018pgm}
\begin{align}
&C_{f, V} \simeq \tilde{g} \cos^2\theta _W Q_f  + g^\prime [Q'_H (T^3_f  - 2 \sin^2\theta _W Q_f ) + (Q'_{f, R}+Q'_{f, L})/2]\, , \nonumber \\
&C_{f, A} \simeq g^\prime [-Q'_H T^3_f + (Q'_{f, R}-Q'_{f, L})/2]\, .
\end{align}
In the following we will simplify the notation and denote 
$Q'_{q_i, R}$ as $Q'_{q_i}$, $Q'_{q_i, L}$ as $Q'_{Q_i}$ with $q$ being $u$ or $d$ for quarks, and $Q'_{\ell_i, R}$ as $Q'_{\ell_i}$, $Q'_{\ell_i, L}$ as $Q'_{L_i}$ with $\ell$ being $e$ or $\nu$ for leptons giving 
\begin{eqnarray}
\label{eq:AxialAndVectorCharges}
  C_{u_i, V} &\simeq&  \frac{2\tilde{g}}{3}\cos^2\theta _W + \frac{g^\prime}{2} \left[Q'_H + Q'_{u_i}+Q'_{Q_i}  - \frac{8}{3}Q'_H \sin^2\theta _W  \right]\, , \nonumber \\
C_{u_i, A} &\simeq& \frac{g^\prime}{2} \left[-Q'_H  + Q'_{u_i}-Q'_{Q_i} \right]\, , \nonumber \\
  C_{d_i, V} &\simeq&  -\frac{\tilde{g}}{3}\cos^2\theta _W + \frac{g^\prime}{2} \left[-Q'_H + Q'_{d_i}+Q'_{Q_i}  + \frac{4}{3}Q'_H \sin^2\theta _W   \right]\, , \nonumber \\
C_{d_i, A} &\simeq& \frac{g^\prime}{2} \left[Q'_H  + Q'_{d_i}-Q'_{Q_i} \right]\,, \nonumber \\
  C_{\nu_i, V} &\simeq&   \frac{g^\prime}{2} \left[Q'_H +Q'_{L_i}  \right]\, , \nonumber \\
C_{\nu_i, A} &\simeq& \frac{g^\prime}{2} \left[-Q'_H  -Q'_{L_i} \right]\, , \nonumber \\
  C_{e_i, V} &\simeq&  -\tilde{g}\cos^2\theta _W + \frac{g^\prime}{2} \left[-Q'_H + Q'_{e_i}+Q'_{L_i}  + 4Q'_H \sin^2\theta _W   \right]\, , \nonumber \\
C_{e_i, A} &\simeq& \frac{g^\prime}{2} \left[Q'_H  + Q'_{e_i}-  Q'_{L_i} \right]\,.
\end{eqnarray}

\subsection{The Higgs Sector}
The scalar potential of the model is given by
\begin{equation}
\label{eq:HM}
V(H,\chi) =  -\mu^2|H|^2 +  \lambda |H|^4 -\mu_\chi^2 |\chi|^2 + \lambda_\chi |\chi|^4 +\kappa  |\chi|^2|H|^2,  
\end{equation}
where $H$ is the SM Higgs doublet and $\kappa$ is the mixing parameter coupling the SM Higgs field to the new scalar $ \chi $. Unlike the SM Higgs sector, this model features two physical Higgs states with two non-vanishing VEVs, $v$ and $v'$ as already mentioned, which in detail are defined as follows:
\bea
\label{eq:vev}
\langle H \rangle = \frac{1}{\sqrt{2}} \left( \begin{tabular}{c} 0 \\ $v$ \end{tabular} \right) \,, \qquad \langle\chi\rangle = \frac{v'}{\sqrt{2}}\,.
\eea
Following Spontaneous Symmetry Breaking (SSB), the physical CP-even Higgs mass eigenstates  ($h_1$ and $h_2$) emerge from the interaction states ($H$,$ \chi $) through an orthogonal rotation:
{\bea
\left( \begin{array}{c} h_1 \\ h_2 \end{array} \right) = \left( \begin{array}{cc} \cos \theta & - \sin \theta \\  \sin \theta & \cos \theta \end{array} \right)  \left( \begin{array}{c} H  \\ \chi \end{array} \right),
\eea 
with the mixing angle $\theta$ constrained to  $- \pi/2 < \theta < \pi/2$. The corresponding mass eigenvalues are given by:
\bea
m_{h_{1,2}}^2 = \lambda v^2 + \lambda\chi  v'^2 \mp \sqrt{\left( \lambda v^2 - \lambda\chi  v'^2\right)^2 + \left( \kappa v v' \right)^2}.
\eea
The mixing angle can be determined from parameters of the scalar potential via 
\bea
\tan 2 \theta = \frac{\kappa v v'}{\lambda v^2 - \lambda\chi  v'^2}.
\label{theta}
\eea}
In the model, $h_2$ is assumed as the SM-like Higgs boson while the exotic boson $h_1$ is lighter and dominantly a singlet-like Higgs state, which has been considered in~\cite{Hicyilmaz:2022owb} as a mediator for possible $Z^\prime$ signatures.

\subsection{The Yukawa sector}
The Yukawa sector of the SM Lagrangian reads:
\be 
- \mathcal{L}_{\rm Yuk.}^{\rm SM} = Y_u \bar{Q} \tilde{H} u_R + Y_d \bar{Q} H d_R + Y_e \bar{L} H e_R \, .
\label{Yukawa}
\ee
Imposing gauge invariance under the $U(1)'$ then gives the following charge relations
\begin{equation}
-Q'_{Q_i} -Q'_H +Q'_{u_i} = -Q'_{Q_i} + Q'_H +Q'_{d_i} = -Q'_{L_i} + Q'_H + Q'_{e_i} = 0.
\label{eq:gauge_invariance}
\end{equation}
As is evident from 
Eq.~(\ref{eq:AxialAndVectorCharges}), 
these conditions eliminate axial-vector $Z'$ couplings for quarks and charged leptons. 
To circumvent this, our model introduces flavour-dependent $U(1)'$ charges to generate crucial axial-vector couplings with nucleons, evading experimental constraints \cite{Feng:2016jff,Feng:2016ysn,NA482:2015wmo}. 
This is achieved through a mechanism where only third-generation fermions obtain masses via SM-like Yukawa terms while the first- and second-generation masses arise from higher-dimensional operators:
\bea
\label{eq:powers}
- \mathcal{L}_{\rm Yukawa} &=& 
\Gamma^{u}_i \left( \frac{\chi^*
}{M} \right)^{n_{i}} \overline{Q}_{L,i}\tilde{H}u_{R,i} 
+ \Gamma^{d}_i\left( \frac{\chi}{M} \right)^{l_{i}} \overline{Q}_{L,i} H d_{R,i} \nonumber \\
&+&\Gamma^{e}_i \left(\frac{\chi}{M} \right)^{m_{i}} \overline{L}_{i} H e_{R,i}+ h.c., \quad i=1,2 \, ,
\eea
where $M$ sets the non-renormalisable scale and we assume that the couplings are flavour diagonal in order to avoid tree level flavour changing neutral currents. Requiring gauge invariance under the $U(1)'$ then gives
\begin{align}
\label{eq:gauge_invariance_gen12}
& -n_{1,2} Q'_{\chi}-Q'_{Q_{1,2}} -Q'_H +Q'_{u_{1,2}}  = 0 \nonumber \\
& l_{1,2} Q'_{\chi}-Q'_{Q_{1,2}} + Q'_H +Q'_{d_{1,2}}  = 0   \nonumber \\
&m_{1,2} Q'_{\chi} -Q'_{L_{1,2}} + Q'_H +Q'_{e_{1,2}}   = 0.
\end{align}
which will generate  non-zero axial couplings when $n_i$, $m_i$ or $l_i\ne0$. In the following we assume flavour-universality of the first two generations of $U(1)'$ quark charges ($Q'_{Q_1} = Q'_{Q_2}$ etc.), 
whereas the lepton charges are fully non-universal. As a consequence we will have $ n_{1} =n_{2} $ and $ l_{1} =l_{2} $.

The $U(1)'$ charges must also satisfy the anomaly cancellation conditions for the fermionic content of the SM and the additional $R$-handed neutrinos:
\begin{align}
\label{eq:anomaly}
& \sum_{i=1}^{3} (2 Q'_{Q_i} - Q'_{u_i} - Q'_{d_i}) = 0 \,,  \\
&  \sum_{i=1}^{3} \, ( 3 Q'_{Q_i} +  Q'_{L_i})  = 0 \,,  \\
& \sum_{i=1}^{3} \left( \frac{Q'_{Q_i}}{6} - \frac{4}{3} Q'_{u_i} - \frac{Q'_{d_i}}{3}  +   \frac{Q'_{{L_i}}}{2} - Q'_{e_i}\right) = 0 \,, 
\label{eq:anomaly3} \\
& \sum_{i=1}^{3} \left( Q_{Q_i}^{\prime 2} - 2 Q_{u_i}^{\prime 2}  + Q_{d_i}^{\prime 2}       - Q_{{L_i}}^{\prime 2}  + Q_{e_i}^{\prime 2}  \right) = 0 \,, \label{eq:anomaly4} \\
& \sum_{i=1}^{3} \left( 6 Q_{Q_i}^{\prime 3}  - 3 Q_{u_i}^{\prime 3}  - 3 Q_{d_i}^{\prime 3}  +  2 Q_{{L_i}}^{\prime 3}  - Q_{e_i}^{\prime 3} \right)  + \sum_{i=1}^{3} Q_{\nu _i}^{\prime 3}    = 0 \,, \label{eq:anomaly5} \\
& \sum_{i=1}^{3} \left( 6 Q'_{Q_i} - 3 Q'_{u_i} - 3 Q'_{d_i}  + 2 Q'_{{L_i}} - Q'_{e_i} \right) + \sum_{i=1}^{3} Q'_{\nu _i}    = 0 .
\label{eq:anomaly6}
\end{align}
The anomaly equations, Eqs.~(\ref{eq:anomaly})--(\ref{eq:anomaly3}),
together with the gauge invariance conditions for the first and second generations, Eq.~(\ref{eq:gauge_invariance_gen12}), as well as the third generation, Eq. (\ref{eq:gauge_invariance}), gives
\begin{eqnarray}
\label{eq:ChiPowers}
n_{1} & = & l_{1} \,, \nonumber  \\
2n_{1} & = & m_{1}+m_{2} \, ,
\end{eqnarray}
which in turn gives the following axial couplings for the first two generations:
\begin{eqnarray}
\label{eq:AxialCharges}
C_{u, A} &\simeq& \frac{g^\prime}{2} \frac{m_{1}+m_{2}}{2} Q'_{\chi}\, , \nonumber \\
C_{d, A} &\simeq& -\frac{g^\prime}{2} \frac{m_{1}+m_{2}}{2} Q'_{\chi} \, , \nonumber \\
C_{e, A} &\simeq& -\frac{g^\prime}{2} m_{1} Q'_{\chi}\, , \nonumber \\
C_{\mu, A} &\simeq& -\frac{g^\prime}{2} m_{2} Q'_{\chi}\,,
\end{eqnarray}
where $(m_{1}+m_{2})/2$ has to be a positive integer with the assumptions made.
Using that $m_{Z'}= Q'_{\chi} g' v'$ then gives $C_{u, A} = \dfrac{m_{1}+m_{2}}{4} \dfrac{m_{Z'}}{v'}$ etc.
From the relations in Eq.~(\ref{eq:AxialCharges}) it is clear that it is not possible, with the assumptions made, to have the axial charges of both the electron and muon vanishing at the same time as the axial charges of the up and down quarks are non-zero. However, this constraint would be removed if the assumption of universality of the first two generation of $U(1)'$ quark charges is lifted. 

In order to solve the remaining anomaly equations we divide them into two groups: one only involing the SM fermions and one also involving the right handed neutrinos. As a first step we use Eq.~(\ref{eq:ChiPowers}) and solve 
Eqs.~(\ref{eq:anomaly})--(\ref{eq:anomaly4})
together with the gauge invariance conditions for first and second generations, Eq. (\ref{eq:gauge_invariance_gen12}), as well as the third generation, Eq. (\ref{eq:gauge_invariance}), giving for the case $m_1 \ne m_2$:
\begin{align}
\label{eq:charges_new}
& Q'_{Q_1} = \dfrac{((m_1+m_2)^2-4m_2^2)Q'_{\chi} + 4(m_1 - m_2) Q'_{e_2} + 4(5m_1- m_2) Q'_{H} +   12 m_1  Q'_{d_3}  + 4m_1 Q'_{e_3} }{12 (m_2 - m_1)} 
  \,,  \nonumber \\
& Q'_{Q_3}=Q'_{d_3}+Q'_{H}   \,,  \nonumber \\
& Q'_{d_1} = Q'_{Q_1} - Q'_{H} - \dfrac{m_1 + m_2}{2}Q'_{\chi} \,,  \nonumber \\
& Q'_{u_1} = Q'_{Q_1} + Q'_{H} + \dfrac{m_1 + m_2}{2}Q'_{\chi} \,,  \nonumber \\
& Q'_{u_3} = Q'_{d_3} + 2 Q'_{H} \,,  \nonumber \\
& Q'_{e_1} = -6 Q'_{Q_1} - 3 Q'_{d_3} - Q'_{e_2} - Q'_{e_3} - 6 Q'_{H} - (m_1+m_2)Q'_{\chi} \,,  \nonumber \\
& Q'_{L_1} =  m_1 Q'_{\chi} + Q'_{e_1} + Q'_{H} \,,  \nonumber \\
& Q'_{L_2} = m_2Q'_{\chi} + Q'_{e_2} + Q'_{H} \,,  \nonumber \\
& Q'_{L_3} = Q'_{e_3} + Q'_{H} \,,  
\end{align}
with $m_1$, $m_2$, $Q'_{\chi}$, $Q'_{d_3}$,  $Q'_{e_2}$, $Q'_{e_3}$ and $Q'_{H}$ being free parameters. 
For the case $m_1 = m_2$ we get the same solutions as above except 
\begin{align}
\label{eq:charges_new_neq}
& Q'_{Q_1} = \dfrac{-2m_2 Q'_{\chi} - Q'_{e_1} - Q'_{e_2} -2  Q'_{H} }{6} \,,  \nonumber \\
& Q'_{e_3} =  - 3 Q'_{d_3} - 4 Q'_{H}  
  \, ,
\end{align}
such that $Q'_{e_1}$ replaces $Q'_{e_3}$ as a free parameter.

The remaining anomaly equations, Eqs.~(\ref{eq:anomaly5})--(\ref{eq:anomaly6}), can then be solved to find $Q'_{\nu_1}$, $Q'_{\nu_2}$ and $Q'_{\nu_3}$ using the methods introduced in~\cite{Rathsman:2019wyk} although in this study we have not done so explicitly and instead solved them numerically.

\subsection{Parameter space explorations and experimental constraints}

To investigate the parameter space of our model we have employed the SPheno~\cite{Porod:2003um,Porod:2011nf,Braathen:2017izn} and SARAH 4.14.3~\cite{Staub:2013tta,Staub:2015kfa} codes. The scanning of the parameter space was performed using the Metropolis-Hastings algorithm, within the ranges specified in Tab.~\ref{paramSP}. We scan integer solutions for $ |Q'|\leq200 $ then normalise by 100 to obtain $ |Q'|<2 $ rational charges. This cutoff balances the need for solution diversity against the unavoidable computational limits.

In addition to the NSI constraints from IceCube \cite{IceCubeCollaboration:2021euf}, some important experimental outcomes have been implemented to our solutions. We first require the SM Higgs boson mass to be within $3$~GeV of its observed value of 125 GeV and its Branching Ratios (BRs), BR$(h_2 \to b\bar{b})$ and BR$(h_2 \to {\rm invisible}~[{\rm i.e., }~h_1 h_1,Z' Z',Z Z'])$ \cite{ParticleDataGroup:2024cfk,ATLAS:2023tkt}\footnote{Wherein $Z\to {\rm neutrinos}$ while the $Z'$ and (light) Higgs state, $h_1$, decay products, $e^+e^-$ and (mainly) $b\bar b$, respectively, do not pass  detector triggers and/or analysis thresholds.}. Then, we import the constraints on the BRs of rare $B$-decays, specifically $ {\rm BR}(B \rightarrow X_{s} \gamma) $, $ {\rm BR}(B_s \rightarrow \mu^+ \mu^-) $ and $ {\rm BR}(B_u\rightarrow\tau \nu_{\tau}) $ \cite{HFLAV:2012imy,LHCb:2012skj,HFLAV:2010pgm}. We have also bounded the value of the $ Z-Z' $ mixing parameter $\theta'$ (see Eq.~(\ref{ThetaPrime})) to be less than a few times $ 10^{-3} $ as a result of EW  Precision Tests (EWPTs)~\cite{DELPHI:1994ufk,Erler:2009jh}. 
\begin{table}[t!]
	\centering
	\begin{tabular}{c|c||c|c}
		\hline
		Parameter  & Scanned range & Parameter      & Scanned range \\
		\hline
		$g'$ & $[10^{-5}, 5\times 10^{-5}]$      & $\lambda$ & $[0.125, 0.132]$ \\
		$\tilde{g}$        & $[-10^{-3}, 10^{-3}]$ & ${\lambda}_{\chi}$ & $[10^{-3}, 10^{-1}]$ \\
		$v'$ & $[0.1, 1]$ TeV  &  $\kappa$  & $[10^{-4}, 10^{-2}]$ \\
		$  Q'_{e_1}/Q'_{e_3}, Q'_{e_2}, Q'_{d_3},Q'_{H},Q'_{\chi}$ & $[-2, 2]$ &$Q'_{\nu_1} $& $0$ \\
        $  m_{1} $ & $[0, 3]$ & $m_{2}$ &  $[0, 3]$ \\
		\hline
	\end{tabular}
	\caption{Scanned parameter space of the theoretical model. Note that $m_1$ and $m_2$ are integers and that if $m_1 \ne m_2$ then $Q'_{e_3}$ is a free parameter which is replaced with $Q'_{e_1}$ if $m_1 = m_2$. }
	\label{paramSP}
\end{table}
In the next part of the numerical analyses, we constrain the parameter space to satisfy the current experimental bounds on the anomalous magnetic moment results for $ (g-2)_{e} $, $ (g-2)_{\mu} $ \cite{Aliberti:2025beg,Muong-2:2025xyk,Morel:2020dww}, the related $Z'$ couplings satisfying the ATOMKI anomaly \cite{Barducci:2022lqd}, the electron beam dump experiment NA64 \cite{NA64:2019auh}, which probes the $Z'$ couplings to electrons, $ C^2_{e,V}+ C^2_{e,A} $, in any possible production and detection via $ e^- + Z \to e^- + Z'[\to e^+ e^-] $ and  neutrino scattering from the TEXONO experiment \cite{TEXONO:2009knm}. {We have also applied additional experimental limits on parity-violating Moller scattering \cite{SLACE158:2005uay,Kahn:2016vjr}, atomic parity violation \cite{Porsev:2009pr,Arcadi:2019uif} and CE$\nu$NS \cite{Denton:2023gat,AristizabalSierra:2022axl,AtzoriCorona:2022moj,Coloma:2022avw} as $  |C_{e,V} \times C_{e,A}| \leq 10^{-8} $,  $ |C_{e,A}| \left| \frac{188}{399}  C_{u,V}  + \frac{211}{399} C_{d,V }\right| \leq 1.8\times10^{-12}  $ and $ \sqrt{\left| C_n C_{\nu_e}\right|} \leq 13.6 \times 10^{-5}$, respectively, Here, $ C_n $ is the $Z'$-neutron coupling defined as $ C_n =  C_{u,V}  + 2 C_{d,V}$. The experimental constraints are summarised in Tab.~\ref{tab:constraints}. }
\begin{table}[t!]
\centering
\begin{tabular}{c|c|c|c}
\hline
Observable & Constraint & Tolerance & Reference(s) \\		
\hline
$m_{h_2}$ & 122 GeV -- 128 GeV & ~ & ~ \\
BR$(h_2 \to b\bar{b})$ & $>0.4$& ~ & \cite{ParticleDataGroup:2024cfk} \\
BR$(h_2 \to {\rm invisible})$ & $<0.17$& $1\sigma$ & \cite{ATLAS:2023tkt} \\
${\rm BR}(B_s \rightarrow \mu^{+} \mu^{-} )$ &
$0.8 \times 10^{-9}$  --
$6.2 \times 10^{-9}$ &
$2\sigma$ & 
\cite{LHCb:2012skj} \\
${\rm BR}(B \rightarrow X_{s} \gamma)$ &
$2.99 \times 10^{-4}$ --
$3.87 \times 10^{-4}$ &
$2\sigma$ & 
\cite{HFLAV:2012imy} \\
$\frac{{\rm BR}(B_u\rightarrow\tau \nu_{\tau})}{{\rm BR}(B_u\rightarrow \tau \nu_{\tau})_{\rm SM}}$ &
0.15  --
2.41 &
$3\sigma$ & 
\cite{HFLAV:2010pgm} \\
$\Delta a_e^{\text{Rb}}$ &
$(4.8 \pm 9.0) \times 10^{-13}$ &
$3\sigma$ & 
\cite{Morel:2020dww} \\
$\Delta a_{\mu}$ &
$(3.8 \pm 6.3)\times 10^{-9}$ &
$3\sigma$ & 
\cite{Aliberti:2025beg,Muong-2:2025xyk} \\
$\epsilon_{ee}^{\oplus }-\epsilon_{\mu\mu}^{\oplus }$ &
$[-2.26, -1.27] \cup [-0.74, 0.32]$ &
~ &
\cite{IceCubeCollaboration:2021euf} \\
$\epsilon_{\tau \tau}^{\oplus }-\epsilon_{\mu\mu}^{\oplus } $ &
$[-0.041,0.042]$ &
~ &
\cite{IceCubeCollaboration:2021euf} \\
$ \sqrt{C^2_{e,V}+ C^2_{e,A}} $ &
$\geq 3.6 \times 10^{-5} \times \sqrt{{\rm BR}(Z' \to e^+ e^-)}$ &
~ &
\cite{NA64:2019auh} \\
$ \sqrt{C_{e,V} \times C_{\nu_e}} $ &
$\leq 3 \times 10^{-4} $ &
~ &
\cite{TEXONO:2009knm} \\
$ |C_{e,V} \times C_{e,A}| $ &
$\leq 10^{-8}  $ &
~ &
\cite{SLACE158:2005uay,Kahn:2016vjr} \\
$ |C_{e,A}| \left| \frac{188}{399}  C_{u,V}  + \frac{211}{399} C_{d,V }\right|  $ &
$\leq 1.8 \times 10^{-12}  $ &
~ &
\cite{Porsev:2009pr,Arcadi:2019uif}  \\
$ \sqrt{|C_{n} \times C_{\nu_e}|} $ &
$\leq 13.6 \times 10^{-5} $ &
~ &
\cite{Denton:2023gat,AristizabalSierra:2022axl,AtzoriCorona:2022moj,Coloma:2022avw}\\
\hline
\end{tabular}
\caption{Summary of the experimental constraints used.}
\label{tab:constraints}
\end{table}

\section{Theory for CE$\nu$NS at spallation sources}\label{Sect3}
Spallation sources, such as the ESS, are not only producing neutrons, but also neutrinos that for example can be used to study the CE$\nu$NS process.
In short, the spallation source produces neutrons by impinging a nuclear target with a high intensity proton beam and the neutrons are produced as the produced nuclei decay into lighter fragments. As part of this process there are also charged pions produced. 

Negatively charged pions $\pi^-$ are absorbed by the nuclei before they can decay, whereas positively charged ones  $\pi^+$ are stopped in the target where they almost exclusively decay into muons and muon neutrinos, $\pi^+ \to \mu^+ \nu_\mu$ with a life time of 26 ns. In turn, the muons decay to electrons and neutrinos, $ \mu^+ \to \bar{\nu}_\mu e^+ \nu_e $, with a life time of 2.2 $\mu$s. Effectively, each $\pi^+$ then gives three neutrinos, $\pi^+ \to \nu_\mu \bar{\nu}_\mu \nu_e  e^+ $ and using that the $\pi^+$ decay at rest, the resulting neutrino flux is isotropic.
In the following, we will concentrate on neutrinos produced at the ESS, which has a long proton pulse and thus we will not distinguish between the prompt $\nu_\mu$ and the  $ \bar{\nu}_\mu \nu_e $ which are more spread out in time.
\begin{figure}[t]
\label{fig_neutrinospectra}
\begin{center}
\includegraphics[width=8cm]{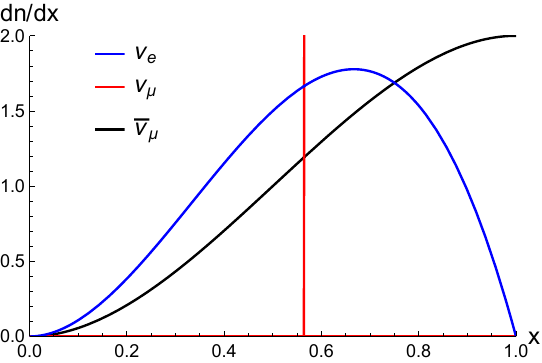}
\end{center}
\caption{Neutrino spectra in the scaled energy variable $x=\dfrac{2E_\nu}{m_\mu}$ from  $\pi^+$ decays at rest normalised to 1.}
\end{figure}
Assuming that the $\pi^+$ as well as the $\mu^+$ decay at rest in the spallation target, the resulting neutrino fluxes can be written in terms of the scaling variable $x=\dfrac{2E_\nu}{m_\mu}$ (with $E_\nu$ being the neutrino energy and $0<x<1$) as
\begin{eqnarray}
 \dfrac{d\Phi_{\nu_e}}{dx} & = & \dfrac{r N_{\mathrm{POT}}}{4\pi L^2}  \,  12x^2(1-x) \label{eq_nue-flux},\\
  \dfrac{d\Phi_{\bar{\nu}_\mu}}{dx} & = & \dfrac{r N_{\mathrm{POT}}}{4\pi L^2}  \,  2x^2(3-2x)   \label{eq_numubar-flux},\\
  \dfrac{d\Phi_{\nu_\mu}}{dx} & = &  \dfrac{r N_{\mathrm{POT}}}{4\pi L^2} \, \delta \left( x- x_0 \right), \label{eq_numu-flux}
\end{eqnarray}
where $r$ is the number of $\pi^+$ produced per proton, $N_{\mathrm{POT}}$ is the number of protons on target, $L$ is the distance travelled by the neutrinos and 
\begin{equation}
x_0 = \dfrac{m_{\pi^+}^2-m_\mu^2}{m_\mu m_{\pi^+}} \approx 0.564 \, 
\end{equation}
is the scaled energy of the muon neutrinos from the two body decay.

The neutrinos produced from the spallation target can be used to study various physics processes. 
In coherent elastic neutrino nucleus scattering they interact with a nuclear target and the nuclear recoil energy is measured. To calculate the nuclear recoil energy spectrum we need the differential cross section for coherent elastic neutrino nucleus scattering of a neutrino ${\nu_\ell}$ from a nucleus with mass $M$ is given by~\cite{Freedman:1973yd}
\begin{eqnarray*}
 \dfrac{d \sigma^{\nu_\ell N}}{dy} & = &  
  \dfrac{G_F^2 m_\mu^2}{4\pi} \left( \dfrac{1}{ 2} - \dfrac{y}{ 2x^2} - \dfrac{ym_\mu}{ 2xM} \right) (Q^{\nu_\ell}_V)^2  + 
  \dfrac{G_F^2 m_\mu^2}{4\pi} \left(  \dfrac{1}{ 2} + \dfrac{y}{ 2x^2} - \dfrac{ym_\mu}{ 2xM} \right) (Q_A^{\nu_\ell})^2,  
\end{eqnarray*}
where $y=E_r/E_r^{\max}$ is the scaled nuclear recoil energy with $E_r^{\max} = \dfrac{m_\mu^2} {2M}$. In addition, with these definitions of $x$ and $y$,  we have\footnote{The inequality $y<x^2$ is valid for $m_\mu \ll M$. At the same time, in this limit the $ \dfrac{ym_\mu}{ 2xM}$ term in the differential cross section does not  contribute to the nuclear recoil spectrum.} $0<y<x^2$.
The $Q^{\nu_\ell}_V$ and  $Q^{\nu_\ell}_A$ term are the vector and axial vector effective charges, respectively,  for the interaction given by
\begin{eqnarray}
\label{eq_weak_charges}
Q_V^{\nu_\ell}  & = &  g_V^{p,{\nu_\ell}} Z F_{V,Z}(y) + g_V^{n,{\nu_\ell}} N F_{V,N}(y), \\
Q_A^{\nu_\ell}  & = &  g_A^{p,{\nu_\ell}}  \langle S_p \rangle  F_{A,Z}(y) + g_A^{n,{\nu_\ell}}   \langle S_n \rangle   F_{A,N}(y),
\end{eqnarray}
where $F_V(y)$ and $F_A(y) $ are the nuclear form factors which depend on the squared momentum transfer $Q^2 = 2M E_r=ym_\mu^2$ and $Z/N$ are the number of protons/neutrons in the nucleus whereas  $\langle S_p \rangle$/$\langle S_n \rangle$ is the corresponding average spin.  
We also use a convention where $F_V(y)$ is normalised such that $F_V(0)=1$. In general, the axial contributions are much smaller than the vector ones. For an  even-even nucleus we even have $F_A(y) =0$. In the following, we will therefore only consider the vector part of the interaction.

\begin{figure}[t]
\begin{center}
\includegraphics[width=8
cm]{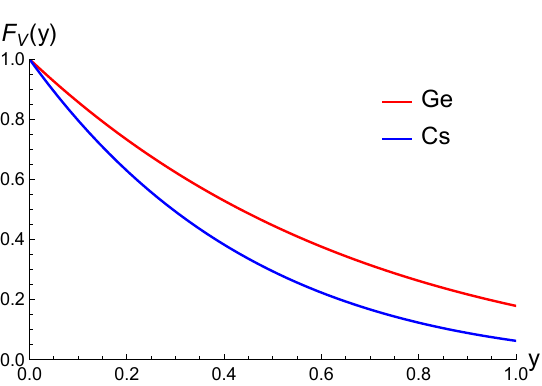} 
\end{center}
\caption{The nuclear form factors $F_{V}(y) =F_{V,Z}(y) =F_{V,N}(y)$ for Germanium ($A=72.6$) and Cesium ($A=133$). } 
\label{fig_formfactor}
\end{figure}

The vector contributions from an additional vector boson exchange adds to the scattering through  
\begin{eqnarray}
\label{eq_weak_couplings}
g_V^{p,{\nu_\ell}} & = &g_{V,\mathrm{SM}}^p + \dfrac{\sqrt{2}C_{{\nu_\ell},V} (2 C_{u,V} +C_{d,V})}{G_F(y m_\mu^2+ m_{Z^\prime}^2)},  \\
g_V^{n,{\nu_\ell}}  & = & g_{V,\mathrm{SM}}^n + \dfrac{\sqrt{2}C_{{\nu_\ell},V} (2 C_{u,V} +C_{d,V})}{G_F(y m_\mu^2+ m_{Z^\prime}^2)},  
\end{eqnarray}
where for the SM contributions we use the Leading Order (LO) values
\begin{eqnarray}
 g_{V,\mathrm{SM}}^p & = & 1-4\sin^2\theta_W\, , \\
 g_{V,\mathrm{SM}}^n & = & -1 \, .
\label{eq_SM_couplings}
\end{eqnarray}
The final pieces needed for calculating the cross section are the form factors $F_{V,Z}(y)$ and $F_{V,N}(y)$. Here we us the Klein-Nystrand model \cite{Klein:1999qj}, assuming $F_{V,Z}(y) = F_{V,N}(y) = F_{V}(y) $, with 
\begin{equation}
F_{V}(y)  = \dfrac{3 \,  j_1(\sqrt{y}m_\mu R_A)}{\sqrt{y}m_\mu R_A}\dfrac{1}{1+a^2ym_\mu^2},
\end{equation}
where $j_1$ is a spherical Bessel function of the first kind, i.e., {$j_1(z)=(\sin z-z\cos z)/z^2$}, $R_A =1.2 A^{1/3} $ fm  and $a=0.7$ fm. The resulting form factors for Germanium ($A=72.6$) and Cesium ($A=133$), respectively, are shown in Fig.~\ref{fig_formfactor}.
From the figure it is clear that the coherence condition, as implemented using the form factor, gives a substantial suppression of larger recoil energies, especially considering that the form factor enters squared in the cross section.

In order to get the nuclear recoil energy spectrum we need to 
integrate the differential cross section for coherent elastic neutrino nucleus scattering over the neutrino energy spectrum, which gives the following nuclear recoil energy spectrum 
\begin{eqnarray}
 \dfrac{d N_r}{dy} & = & \dfrac{ m_{\rm target}}{M_{\rm A}} N_{\rm A}  \sum_{\nu_\ell} \int_{x_{\min}}^{x_{\max}}  \dfrac{d \sigma^{\nu_\ell N}}{dy}  \dfrac{d \Phi_{\nu_\ell}}{dx} dx, 
\end{eqnarray}
where $x_{\min} =   \sqrt{y}$ and $x_{\max}=1$
and we have also included a factor $ \dfrac{ m_{\rm target}}{M_{\rm A}} N_{\rm A} $ for the number of nuclei in the target such that $N_r$ is the number of recoils. 
Inserting the neutrino fluxes from Eqs.~(\ref{eq_nue-flux})--(\ref{eq_numu-flux}) and integrating over the scaled neutrino energy $x$ gives
\begin{eqnarray}
 \dfrac{d N_r}{dy} & = &  
 \dfrac{r N_{\mathrm{POT}}}{4\pi L^2}   \,  
 \dfrac{G_F^2 m_\mu^2}{4\pi} \,
\dfrac{ m_{\rm target}}{M_{\rm A}} N_{\rm A}  
 \sum_{\nu_\ell}
 \dfrac{dn_{\nu_\ell}}{dy} (Q_V^{\nu_\ell})^2,
\label{eq:recoil_spec}
\end{eqnarray}
with
\begin{eqnarray}
 \dfrac{dn_{\nu_e}}{dy} & = &   \frac{1}{2} - 3 y +4 y^{3/2}  -  \frac{3}{2} y^2\, , \\
 \dfrac{dn_{\nu_\mu}}{dy} & = &    \frac{1}{2} - 2 y +2 y^{3/2}  - \frac{1}{2} y^2 
 + \left(\dfrac{ 1}{2}- \dfrac{ y}{2x_0^2} \right) \Theta\left(1-\dfrac{y}{x_0^2} \right) \, .
\end{eqnarray}
Here, the last term with the Heaviside $ \Theta$-function comes from the mono-energetic muon neutrinos which thus only contribute for $y < x_0^2 \approx 0.318 $. 

For clarity, we rewrite the nuclear recoil energy spectrum using Eqs.(\ref{eq_weak_charges})--(\ref{eq_SM_couplings}) as
\begin{eqnarray}
\label{eq_recoilspec}
\dfrac{dN_r}{dy} & = & \dfrac{r N_{\rm POT}}{4 \pi L^2} \dfrac{1}{2\pi m_\mu^2} [ N - (1-4\sin^2\theta_W)Z]^2  \dfrac{ m_{\rm target}}{M_{\rm A}} N_{\rm A} \, [F_V(y)]^2 \\ \nonumber 
&& \left\{ \left(  \dfrac{G_Fm_\mu^2}{\sqrt{2} }  - \dfrac{C^{\nu_e}_{\rm eff}  }{ y + m_{Z^\prime}^2/m_\mu^2}   \right)^2 \dfrac{dn_{\nu_e}}{dy}
+ \left( \dfrac{G_Fm_\mu^2}{\sqrt{2} } - \dfrac{C^{\nu_\mu}_{\rm eff}  }{ y + m_{Z^\prime}^2/m_\mu^2}   \right)^2 \dfrac{dn_{\nu_\mu}}{dy} 
\right\},
\end{eqnarray}
where
\begin{eqnarray}
\label{eq_EffectiveNeutrinoCoupling}
C^{\nu_{e/\mu}}_{\rm eff} &=&\dfrac{C_{{\nu_{e/\mu}},V} [ N(C_{u,V} + 2 C_{d,V})+ Z(2 C_{u,V} +C_{d,V})] }{[ N - (1-4\sin^2\theta_W)Z]} 
\end{eqnarray}
are the effective neutrino nucleus couplings for light $Z^\prime$ exchange. We also note that, apart from the form factor, the shape of the nuclear recoil energy spectrum is universal for all nuclei when expressed in terms of the scaled nuclear recoil energy, $y$.
Finally, for reference, the $B-L$ model can be obtained by neglecting the kinetic and $Z-Z^\prime$ mixing giving $C_{f,V}=g' Q'_f$, where, with our conventions in Eq.~(\ref{VAcouplings}), $Q^\prime_u = (Q^\prime_{u_1} + Q^\prime_{Q_1})/2  = 1/3$, $Q^\prime_d =(Q^\prime_{d_1} + Q^\prime_{Q_1})/2  = 1/3$,  $Q^\prime_{\nu_e}=Q^\prime_{L_1}/2=-1$ and $Q^\prime_{\nu_\mu}=Q^\prime_{L_2}/2=-1$. 

\begin{table}
\centering
\begin{tabular}{c c c c c c }
\hline\hline
Parameter         	& BM1 		    	& BM2 				    & BM3 				     & BM4 			& $B-L$ 			\\ [0.5ex] 
\hline
$g^\prime$         	& 1.06$\times 10^{-5}$ & 1.87$\times 10^{-5}$  & 4.34$\times 10^{-5}$   &2.25$\times 10^{-5}$ &5.0$\times 10^{-5}$   \\
$m_{Z^\prime}$ [MeV] & 17.3		    	& 17.2 				    & 17.1  				 & 17.4 			& 17.0 			 \\
$Q^\prime_{u_1}$ 	& 3.76 			    &$-0.32$				    & 0.80 				     & 0.36			    &  		1/3		\\
$Q^\prime_{d_1}$ 	&-$1.88$  			& 0.16  			    &$-0.40$ 				     &$-0.18$  			&   	1/3		\\
$Q^\prime_{Q_1}$ 	 &0.94 			    &$-0.08$ 				    & 0.20 				     & 0.09  			& 		1/3		 \\
$Q^\prime_{L_1}$  	&$-1.92$  			&$-0.06$  				&$-0.60$ 				     & 0.09  			& $-2$ 				\\
$Q^\prime_{L_2}$ 	&$-3.68$  			& 0.54  				&$-0.60$  				 & $-0.63$ 			& $-2$  			\\  
$C_{u,V}$  			& $-2.73\times 10^{-4}$ & $-2.59\times 10^{-4}$ & $-2.75\times 10^{-4}$ & $-1.94\times 10^{-4}$ & 1.67$\times 10^{-5}$ \\
$C_{d,V}$  			&  1.32$\times 10^{-4}$ &  1.32$\times 10^{-4}$ &  1.37$\times 10^{-4}$ &  9.30$\times 10^{-5}$  & 1.67$\times 10^{-5}$ \\
$C_{{\nu_e},V}$  	& 1.39$\times 10^{-5}$ &  $-8.40\times 10^{-6}$     &  0                     & 1.22$\times 10^{-5}$  & $-5.00\times 10^{-5}$ \\
$C_{{\nu_\mu},V}$ 	& 4.58$\times 10^{-6}$ & $-2.80\times 10^{-6}$  &  0                      &  4.06$\times 10^{-6}$  & $-5.00\times 10^{-5}$ \\  
$C_{\rm eff}^{\nu_e}$  &  $-0.56\times 10^{-8}$ &  $-0.30\times 10^{-8}$ &  0.73$\times 10^{-8}$ & 0.19$\times 10^{-8}$  & $-0.46\times 10^{-8}$ \\
$C_{\rm eff}^{\nu_\mu}$ & $-0.23\times 10^{-8}$ & $-0.48\times 10^{-8}$ & 0.73$\times 10^{-8}$ & 0.38$\times 10^{-8}$  & $-0.46\times 10^{-8}$ \\  [1ex]
\hline
\end{tabular}
\caption{Parameter values defining the four benchmark and $B-L$ models considered.}
\label{table_benchmarks}
\end{table}

In order to illustrate what the nuclear recoils spectrum could look like, we have defined four different benchmark points in the $X17$ model (BM1--4), given in Tab.~\ref{table_benchmarks},  that are allowed by current data. The table also gives the corresponding couplings for the $B-L$ model with $g' = 5.0\times 10^{-5} $, which is the current limit for a mass $m_{Z^\prime} = 17$ MeV.  For numerical comparison we also recall that ${G_Fm_\mu^2}/{\sqrt{2} }=9.3\times10^{-8}$.

\begin{figure}[ht]
\begin{center}
\includegraphics[width=10cm]{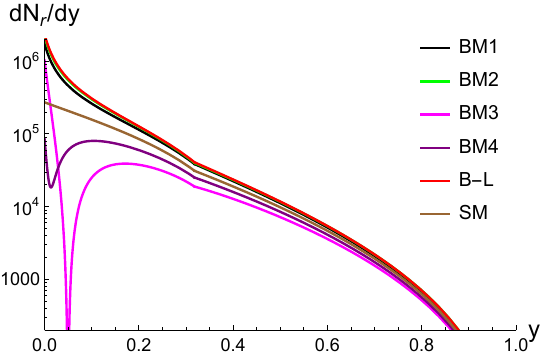} 
\end{center}
\caption{The recoil spectra in the SM and four benchmark models given in Tab.~\ref{table_benchmarks} as well as the $B-L$ model with $g_{Z^\prime} = 5\times 10^{-5}$.} 
\label{fig_benchmarks}
\end{figure}

The resulting recoil spectra are shown in Fig. \ref{fig_benchmarks}. First of all, we note the bump in the spectrum just below $y\approx0.32$ which arises from the mono-energetic muon neutrinos. From the figure we also see that the shape of the spectrum can vary a lot depending on the signs of the effective neutrino couplings and the resulting positive or negative interference between $Z$ and $Z^\prime$ exchange. At the same time, the dependence of the  $Z^\prime$ exchange on the scaled recoil energy $y$ leads to larger effects at small recoil energies. Finally, there can also be specific charge assignments, as in BM3 with $C_{\rm eff}^{\nu_e}=C_{\rm eff}^{\nu_\mu} >0 $, where there is a specific value of $y$ for which the recoil spectrum vanishes.

\begin{figure}[t]
\begin{center}
\includegraphics[width=7.5cm]{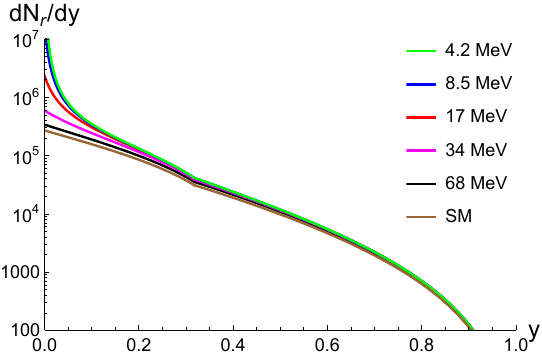} 
\includegraphics[width=7.5cm]{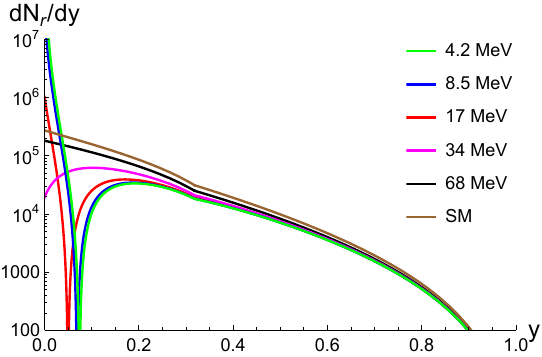} 
\end{center}
\caption{The recoil spectrum in the $B-L$ model (left) with $g_{Z^\prime} = 5\times 10^{-5}$ and the $Z^\prime$ model for BM3 (right) using different  $m_{Z^\prime}$ as indicated in the figure compared to the SM.} 
\label{fig_massdep}
\end{figure}

As it happens, the $Z^\prime$ mass of interest has more or less the optimal value $m_{Z^\prime}^2/m_\mu^2 \approx 0.03 $ to see the effect of the screening of the $Z^\prime$ propagator by the $Z^\prime$ mass. To illustrate this point further, Fig.  \ref{fig_massdep} shows how the recoil spectrum would look in the $B-L$ model with $g_{Z^\prime} = 5\times 10^{-5}$ as well as in the $X17$ model with BM3 couplings for different $Z^\prime$ masses. As can be seen from the figure, for smaller masses the differences become smaller and similarly for larger masses the difference to the SM is much smaller, essentially only affecting the normalisation and not the shape. 

\section{Experimental considerations}\label{Sect4}

In this section we discuss the experimental considerations  connected to foreseen CE$\nu$NS at the ESS in more detail.

\subsection{Detector effects}
In order to get a more realistic picture of what the experimental signal would look like in an CE$\nu$NS experiment, we take into account the following two main effects: conversion of the nuclear recoil energy to ionisation energy and smearing of the ionisation energy due to the detector resolution. 

\begin{figure}[t]
\begin{center}
\includegraphics[width=8cm]{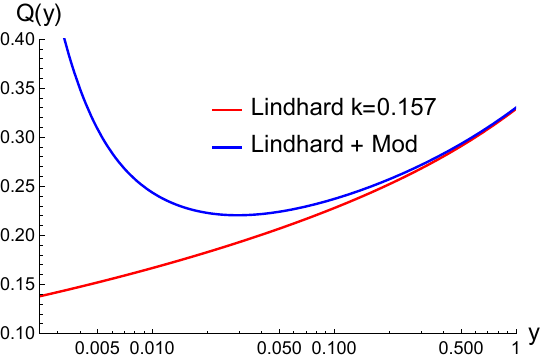} 
\end{center}
\caption{Quenching factor for Germanium. } 
\label{fig_qfactor}
\end{figure}

For the conversion of the nuclear recoil energy to ionisation energy we use the Lindhard model \cite{Scholz:2016qos} to describe the quenching factor,
\begin{equation}
    y_{\rm ion}  = Q(y)y,
\end{equation}
with  
\begin{equation}
    Q(y) = \dfrac{kg(y)}{1 + kg(y)}
\end{equation}
and
$g(y) = 3(C_Z y)^{0.15} + 0.7(C_Z y)^{0.60}  + C_Z y $, where $C_Z = E_r^{\rm max} \times 11.5/Z^{7/3}$ and k = 0.157. The resulting quenching factor is shown 
in Fig. \ref{fig_qfactor} together with a modified version used to span the uncertainties in the measurements of the quenching factor at low nuclear recoil energies \cite{Collar:2021fcl}.

\begin{figure}[t]
\begin{center}
\includegraphics[width=7.5cm]{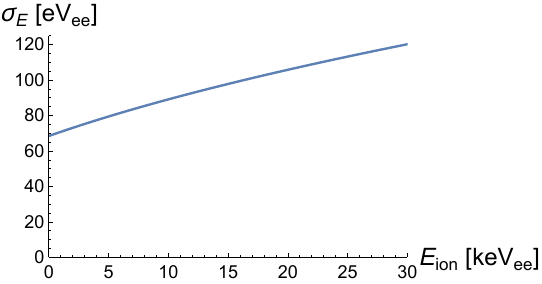}  
\vspace*{0.1cm}
\includegraphics[width=7.5cm]{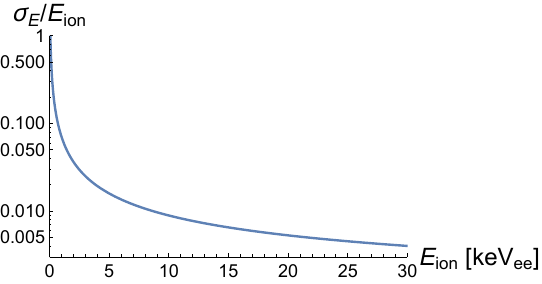} 
\end{center}
\caption{The energy dependence of the resolution and the relative error used to model the Germanium detector. } 
\label{fig_sigma}
\end{figure}

In addition to the quenching factor we also take into account the finite detector resolution, which gives a further smearing of the signal. In terms of the 
reconstructed energy, which we denote as $y_{\rm rec}$, and the ionisation energy, $y_{\rm ion}$ the resolution is given by
\begin{equation}
R(y_{\rm rec},y_{\rm ion})=\dfrac{1}{\sigma \sqrt{2\pi}} \exp\left[ - \dfrac{(y_{\rm rec} - y_{\rm ion})^2}{2\sigma^2 }\right].
\end{equation}
Here, we follow \cite{Coloma:2022avw} and use an energy dependent width
\begin{equation}
\sigma_E^2 =  \sigma_n^2 + E_{\rm ion} \eta F,  
\end{equation}
where $\sigma_n=68.5$ eV, $\eta=2.96$ eV and $F=0.11$, which we convert to a dimensionless width $\sigma=\sigma_E/E_r^{\max}$ giving
\begin{equation}
\sigma  = 0.000830 \sqrt{1 + 5.73 y_{\rm ion} }.
\end{equation}
For illustration, the resulting resolution $\sigma_E$ and relative error
$\dfrac{\sigma_E}{E_{\rm ion}}$ are shown in Fig. \ref{fig_sigma} as a function of the ionisation energy $E_{\rm ion} =y_{\rm ion}  E_r^{\max}$.

Finally, we apply an energy threshold of the detector of $0.200$ keV$_{ee}$ giving a lower limit  $y_{\rm rec} > \dfrac{0.200  \,{\rm keV} }{ 82.5  \,{\rm keV}} = 0.00242$.

With these considerations the recoil spectrum that would be measured is obtained as
\begin{eqnarray}
\dfrac{dN_r}{dy_{\rm rec}} & = &  \int_0^1 R\left(y_{\rm rec},Q(y)y\right) \dfrac{dN_r}{dy}  dy \, .
\end{eqnarray}

\subsection{Normalisation factors}
To get the number of events we also need the various numerical factors in Eq.~(\ref{eq:recoil_spec}) as follows.
The flux factor  
\begin{equation}
\dfrac{r N_{\mathrm{POT}}}{4\pi L^2} 
\end{equation}
depends on the yield $r$, the number of protons on target $N_{\mathrm{POT}}$ and the distance from target to detector $L$.
The yield $r$ depends strongly on the proton energy: for $E_p = 940$ MeV, which is similar to the energy expected at ESS by the end of 2027, we use $r=0.08$  whereas for $E_p = 2.0$ GeV, which is the design goal for ESS, we use $r=0.3$~\cite{Baxter:2019mcx}. 
Realistic placements of a CE$\nu$NS detector at ESS gives $L$ in the range 15 to 25 m and for concreteness we consider ``Near" and ``Far" scenarios with $L=15$ m and 25 m, respectively. 

The number of protons on target  depends on the power ($P$) and the proton energy ($E_p$). In detail $N_{\mathrm{POT}}/s = I /e = P /(eE_p)$ (with $e$ being the elementary charge). Here we also consider two scenarios, which we call ``Low" and ``High". For the ``Low" scenario we assume a proton energy of 840 MeV and power of 800 kW and for the ``High" scenario we use a proton energy of 2.0 GeV and power of 5.0 MW, which gives $N_{\mathrm{POT}}/s =  5.9 \times 10^{15}$ and $N_{\mathrm{POT}}/s =  1.6 \times 10^{16}$, respectively.
With 5000 hours of running time per year thus gives  $N_{\mathrm{POT}} =  1.1 \times 10^{23}$ and $N_{\mathrm{POT}} =  2.8 \times 10^{23}$ per year, respectively. For the ``Low" scenario we assume one year of running time whereas for the   ``High" scenario we assume 5 years. 

The resulting flux factors for the different setups are summarised in Tab.~\ref{table_fluxfactor}.
In addition to the neutrino flux the number of recoils also depends on the number of target nuclei, which is given by the target factor  
\begin{equation} \dfrac{ m_{\rm target}}{M_{\rm Ge}} N_{\rm A} = 1.66 \times 10^{26}\end{equation}
for Germanium using the values $m_{\rm target} = 20$ kg, $M_{\rm Ge}=72.6$ kg/kmol,  and $N_{\rm A}=6.022 \times 10^{26}$/kmol. The flux factor and target factor together gives a luminosity factor
\begin{equation}{\cal L} = \dfrac{r N_{\mathrm{POT}}}{4\pi L^2}  \, \dfrac{ m_{\rm target}}{M_{\rm Ge}} N_{\rm A} \end{equation}
which is also given in Tab.~\ref{table_fluxfactor}.

\begin{table}
\centering
\begin{tabular}{| l | c c | c c | }
\hline
                         &\multicolumn{2}{c|}{Near}   &\multicolumn{2}{c|}{Far}  \\ [0.5ex] 
Parameter                                          & Low            &High  & Low               &High  \\ [0.5ex] 
\hline 
Beam energy [GeV]                &   0.84             &    2.0&   0.84           &    2.0     \\
Yeild $r$                                             &    0.08              & 0.3   &    0.08            & 0.3  \\
Power [MW]                                        &    0.80          & 5.0  &    0.80            & 5.0   \\
$N_{\mathrm{POT}}$  [year$^{-1}$]      &  $1.1\times10^{23}$  & $2.8\times10^{23}$ &  $1.1\times10^{23}$ & $2.8\times10^{23}$    \\
Time in years                                         &    1                                                  & 5                             &    1                                                & 5  \\
Distance $L$    [m]                                &    15                                                &  15                          &    25                                                   &25   \\
Flux factor   [GeV$^{2}$]                     & $1.2\times10^{-13}$  & $5.8\times10^{-12}$   & $4.4\times10^{-14}$ & $2.1\times10^{-12}$    \\[0.5ex]
\hline
Luminosity factor   ${\cal L}$                        & $2.0\times10^{13}$  & $9.6\times10^{14}$  & $7.2\times10^{12}$ & $3.4\times10^{14}$    \\[1ex]
\hline
\end{tabular}
\caption{Parameter values defining the flux factor $\frac{r N_{\mathrm{POT}}}{4\pi L^2} $ considered for different scenarios assuming 5000 hours per year of beam on target. The luminosity factor is the flux factor times the target factor $ \dfrac{ m_{\rm target}}{M_{\rm Ge}} N_{\rm A}$ assuming a 20 kg Germanium target.}
\label{table_fluxfactor}
\end{table}

Finally, to get the number of events we also need the overall normalisation factor for the recoil spectrum, 
\begin{equation} \dfrac{G_F^2 m_\mu^2}{4\pi}  \left(N- (1-4\sin^2\theta_W)Z   \right)^2 = 1.88 \times 10^{-10}  {\rm GeV}^{-2},\end{equation}
for Germanium, using the values $N=40.6$ and $Z=32$ together with the low energy value $\sin^2\theta_W=0.24$.

\subsection{Backgrounds and uncertainties}

\begin{figure}[t]
\begin{center}
\includegraphics[width=7.5cm]{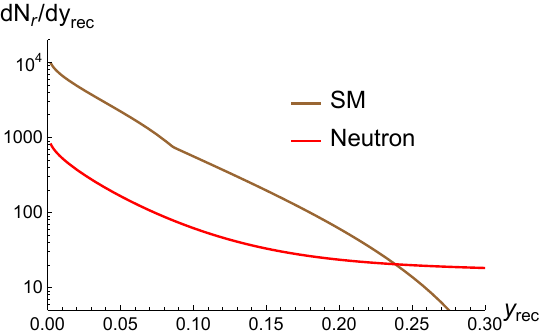} 
\vspace*{0.1cm}
\includegraphics[width=7.5cm]{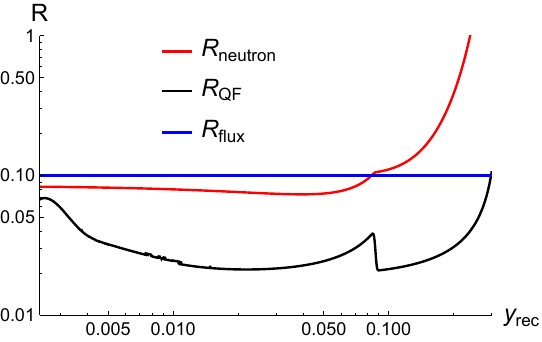} 
\end{center}
\caption{Left: the simulated neutron background from \cite{Lewis:2023sbl} after rescaling from CsI to Ge and including the quenching factor giving the induced recoil spectrum 24 m from the ESS target using a 50 cm polyethylene neutron moderator. Right: the estimated systematic uncertainties in the LowFar scenario.} 
\label{fig_sys}
\end{figure}

Apart from the irreducible SM background we also take into account the neutron induced recoil spectra using the simulations from  \cite{Lewis:2023sbl} after rescaling from CsI to Ge and including the quenching factor. The resulting neutron induced spectrum is shown together with the SM in Fig.~\ref{fig_sys}.

From the neutron induced background we calculate a systematic uncertainty on the SM recoil spectrum 
from the respective spectra  shown in  Fig.~\ref{fig_sys} 
by taking the ratio giving an uncertainty factor
\begin{equation}R_{\rm neutron} = \left.\dfrac{dN_r}{dy_{\rm rec}}\right|_{\rm neutron} / \left.\dfrac{dN_r}{dy_{\rm rec}}\right|_{\rm SM} \, .\end{equation} 

We also take into account the systematic uncertainty arising from the uncertainties in the quenching factor at small recoil energies by defining a factor
 $ R_{\rm QF}$, which is obtained by integrating the theoretical SM spectrum $\dfrac{dN_r}{dy}$ using the two different quenching factors shown in Fig.~\ref{fig_qfactor} and using the relative difference 
 \begin{equation}
 R_{\rm QF} = \left(\left.\dfrac{dN_r}{dy_{\rm rec}}\right|_{\rm QF2} -\left.\dfrac{dN_r}{dy_{\rm rec}}\right|_{\rm QF1}\right)/ \left.\dfrac{dN_r}{dy_{\rm rec}}\right|_{\rm QF1} 
 \end{equation}
 integrated over the respective bins as an estimate. The resulting relative difference is shown in Fig.~\ref{fig_sys}.

The uncertainty in the neutrino flux coming from the pion yield $r$,  for which we use $ R_{\rm flux} = 0.10$, is assumed to be constant. 

We thus consider the systematic uncertainty on the SM prediction from the quenching factor and the neutrino flux, which are both irreducible, as well as the neutron induced background, which is reducible. Being conservative we add these three systematic uncertainties linearly and define a total systematic uncertainty
\begin{equation}
R_{\rm sys} = R_{\rm QF}  + R_{\rm flux} +   R_{\rm neutron}.
\end{equation}
All the systematic uncertainties considered are summarised in Fig.~\ref{fig_sys}.

\section{Discovery reach}\label{Sect5}

\begin{table}
\centering
\begin{tabular}{| c | c c | c c | }
\hline
                         &\multicolumn{2}{c|}{Bin limits}   &\multicolumn{2}{c|}{$N_{\rm SM}$}  \\ [0.5ex] 
Bin nr & $y_{\rm rec}^{\rm low}$ & $y_{\rm rec}^{\rm high}$ & LowFar & HighNear  \\ [0.5ex] 
\hline 
1 & 0.0024 & 0.0080 & 4.61$\times10^{1}$ & 6.13$\times10^{3}$\\
2 & 0.0080 & 0.0150 & 4.46$\times10^{1}$ & 5.94$\times10^{3}$ \\
3 & 0.0150 & 0.0230 & 4.05$\times10^{1}$ & 5.38$\times10^{3}$  \\
4 & 0.0230 & 0.0330 & 3.95$\times10^{1}$ & 5.26$\times10^{3}$ \\
5 & 0.0330 & 0.0450 & 3.57$\times10^{1}$ & 4.75$\times10^{3}$ \\
6 & 0.0450 & 0.0630 & 3.64$\times10^{1}$ & 4.84$\times10^{3}$ \\
7 & 0.0630 & 0.1000 & 3.50$\times10^{1}$ & 4.65$\times10^{3}$ \\
8 & 0.1000 & 0.3000 & 2.53$\times10^{1}$ & 3.37$\times10^{3}$ \\
\hline
\end{tabular}
\caption{Binning used and the number of events expected for each bin in the SM.}
\label{table_binning}
\end{table}

In order to investigate the potential discovery reach of an CE$\nu$NS experiment at ESS we start by exploring the nuclear  recoil spectrum in the LowFar scenario. The spectrum is divided into a number of bins with similar number of events in the SM as given in Tab.~\ref{table_binning}.
For each bin, the systematic uncertainties that we have estimated in section \ref{Sect4} are obtained by integrating over the respective bins and the statisical uncertainties are estimated as $\sqrt{N_{\rm SM}}$.  For the HighNear scenario, we further assume that the systematic errors can be reduced with a factor of 5, 
for example by making a dedicated measurement of the neutrino flux using a heavy water detector.

\begin{figure}[t]
\begin{center}
\includegraphics[width=7.5cm]{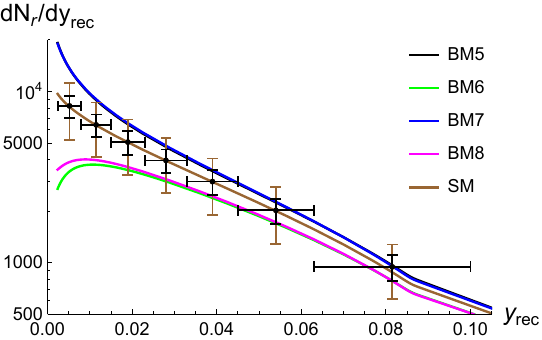} 
\vspace*{0.1cm}
\includegraphics[width=7.5cm]{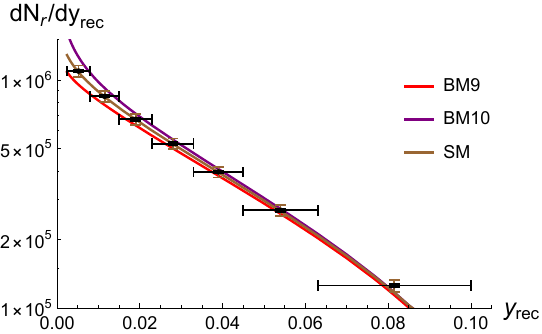} 
\end{center}
\caption{Nuclear recoil spectra for the SM with the projected statistical and systematic errors together with different benchmark points for the $Z^\prime$ model. Black error bars indicate statistical errors and brown statistical and systematic errors added linearly. The left and right plots show the LowFar and HighNear scenarios, respectively.} 
\label{fig_SMEventsPlot}
\end{figure}

The resulting distribution of events $dN/dy_{\rm rec}$ for the SM in the LowFar and HighNear scenarios are shown in Fig.~\ref{fig_SMEventsPlot} together with the statistical and systematic uncertainties. The plots also show the recoil spectra for another four (BM5-BM8) benchmark models which would be possible to exclude at the 95\% Confidence Level (CL) in the LowFar scenario and correspondingly two benchmark models (BM9-BM10) in the HighNear scenario. 
To calculate the signal strength we add the systematic and statistical uncertainties linearly and use the $\chi^2$-function
\begin{equation}
\chi^2 = \sum_{\rm bins} \left(\dfrac{N_{\rm S}}{\sqrt{N_{\rm SM}}+R_{\rm sys} N_{\rm SM}} \right)^2 = \sum_{\rm bins} \left(\dfrac{\sigma_{{\rm SM}+Z^\prime}-\sigma_{\rm SM}}{\sqrt{\sigma_{\rm SM}/\cal{L}}+R_{\rm sys} \sigma_{\rm SM}}  \right)^2
\end{equation}
where we have used that $N_{\rm SM} = \cal{L}\sigma_{\rm SM}$. The  95\% CL exclusion limit then corresponds to $\chi^2 > 6.0$.

From the spectra it is clear that the largest differences are at small recoil energies, which arise from the propagator enhancement, $
C^{\nu_\ell}_{\rm eff}/( y + m_{Z^\prime}^2/m_\mu^2)  $ for $Z^\prime$ exchange as shown in Eq.~(\ref{eq_recoilspec}). Therefore it is important that the detector threshold is as low as possible to maximise the discrimination power. In addition, we recall that for large $C_{\rm eff}^{\nu_\ell} > 0 $, there may also be dips in the spectrum from the negative interference of the $Z^\prime$ with the SM $Z$.

\begin{figure}[t]
\begin{center}
\includegraphics[width=\textwidth]{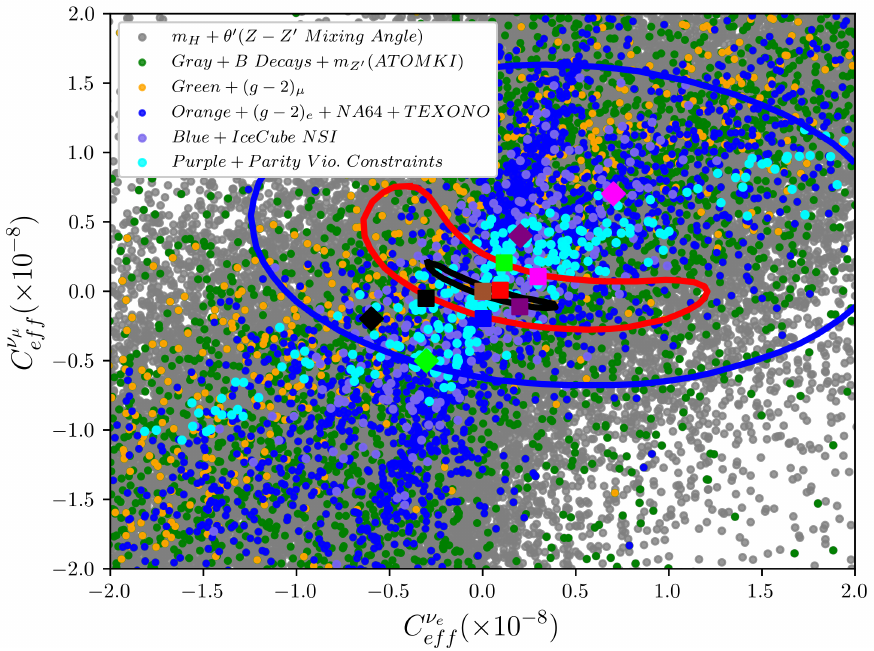}
\vspace*{-0.8cm}
\end{center}
\caption{The currently allowed parameter space from various constraints as indicated in the figure  together with curves showing current (blue) and projected (red/black in LowFar/HighNear scenario, respectively)  95 \% exclusion regions from CE$\nu$NS. The coloured squares are the benchmark points (BM5-BM10) in Fig.~\ref{fig_SMEventsPlot} using the same colour coding and the coloured diamonds are the benchmark points (BM1-BM4) in  Fig.~\ref{fig_benchmarks} again using the same colour coding.} 
\label{fig_chi2}
\end{figure}

Next we consider the discovery reach more generally by scanning over the charges as outlined in section \ref{Sect2} and taking into account the various constraints also listed there. For each set of charges we  calculate the effective neutrino couplings $(C^{\nu_{e}}_{\rm eff},C^{\nu_{\mu}}_{\rm eff})$ and then the corresponding point is given a colour depending on which constraints that it passes as listed in  Fig.~\ref{fig_chi2}. The figure also includes three different contours corresponding to an estimate of the currently excluded points from CE$\nu$NS, as well as points that can be excluded at 95\% CL in the LowFar and  HighNear scenarios, respectively.

From the figure it is clear that, even though the available parameter space in the models probed is already constrained by other experiments, the remaining parameter space can, to a large extent, be covered already in the LowFar scenario at the ESS while the remaining parts would  essentially be covered in the HighNear scenario, except for a very small region close to the SM limit. We also note that, in all scenarios considered, the discovery reach for  $ C^{\nu_{\mu}}_{\rm eff}$ is much larger than for $C^{\nu_{e}}_{\rm eff}$, which just reflects the fact that the total $\nu_{\mu}$ flux is twice as large and also that the $\nu_{\mu}$'s  have higher energy than the $\nu_e$'s.

\section{Conclusions}\label{Sect6}
In summary, spurred by the results obtained by the ATOMKI experiment, 
wherein a more than $5\sigma$ deviation from the SM predictions has been found in the $e^+e^-$ conversion rate of photons produced 
by nuclear transitions   of $^8$Be, $^4$He and $^{12}$C isotopes, generally interpreted as a light $Z'$ with vector-axial coupling and a $\approx17$ MeV mass (hence, aptly named $X17$), we studied the scope of potential CE$\nu$NS experiments, to be hosted at the ESS, in testing this theoretical explanation. Herein, the signal would emerge as not only a modification of the inclusive cross section for $\nu$-nucleus scattering but also of the shape of the nuclear recoil spectrum (with respect to the effects induced by the SM gauge boson $Z$).

In order to do so, we have
embedded the $X17$ within a specific BSM scenario, i.e., a family-dependent $U(1)'$ extension of the SM, which spontaneous breaking introduces a purportedly light vector boson in the required $X17$ guise, in order to have a viable (but not necessarily unique) theoretical framework enabling us to make quantitative predictions for   CE$\nu$NS observables. Upon accounting for assumed statistics and projected systematic errors (from uncertainties due to the neutrino
flux, quenching factor and neutron induced background), we quantified 
the sensitivity that an ESS facility should afford to have in this respect.  

Our overarching conclusion is that a dedicated CE$\nu$NS experiment at the ESS would be able to improve current limits on such a light $Z'$  already within one year of running and ultimately cover a 
substantial part of the relevant parameter space thanks to the significant  statistics foreseen at such a facility. In fact, such an experiment would also be able to discover the $U(1)'$ solution pursued here as an explanation of the ATOMKI anomaly, if realised in Nature. More concretely, the limits on the effective neutrino-nucleus coupling from $Z^\prime$ exchange $(C^{\nu_{e}}_{\rm eff},C^{\nu_{\mu}}_{\rm eff})$, could be reduced from the present ranges ${([-1.2,2.2],[-0.7,1.6])}$, with a small correlation between the two, to ${([-0.3,0.4],[-0.2,0.3]})$, with a strong anti-correlation between these only leaving a small sliver-like region that cannot be probed at the ESS.

\section*{Acknowledgments}
SM is supported in part through the NExT Institute and STFC Consolidated Grant ST/X000583/1. The work of YH is supported by Balikesir University Scientific Research Projects with Grant No. BAP-2022/083. YH thanks SM, JR and Rikard Enberg for their hospitality during a visit to Uppsala, which finalized this work, under the auspices of  Short Term Scientific Mission (STSM) via COST Action CA21106 “COSMIC WISPers”.

\bibliographystyle{JHEP}  


\end{document}